\documentclass[paper,notoc,nohyper]{JHEP3}
\usepackage{amsmath,amssymb,amsfonts}
\usepackage{graphicx}
\usepackage{longtable}
\usepackage{listings}
\usepackage{amsmath}
\usepackage{slashed}

\graphicspath{{pics/}}

\newcommand{\GOSAM}{{\textsc{Go\-Sam}}}

\newcommand{\SAMURAI}{{\textsc{Sa\-mu\-rai}}}
\newcommand{\GOLEMVC}{{\texttt{Go\-lem95C}}}

\newcommand{\FEYNRULES}{{\texttt{Feyn\-Rules}}}
\newcommand{\UFO}{{\texttt{UFO}}}

\newcommand{\SOFTSUSY}{{\textsc{SOFTSUSY}}}
\newcommand{\SDECAY}{{\textsc{SDECAY}}}

\newcommand{\SUSYHIT}{{\textsc{SUSYHIT}}}
\newcommand{\SLHA}{{\textsc{SLHA}}}
\newcommand{\FEYNARTS}{{\texttt{FeynArts}}}
\newcommand{\FORMCALC}{{\texttt{FormCalc}}}

\newcommand{\bea}{\begin{eqnarray*}}
\newcommand{\eea}{\end{eqnarray*}\noindent}
\newcommand{\beq}{\begin{equation}}
\newcommand{\eeq}{\end{equation}\noindent}
\newcommand{\bcen}{\begin{center}}
\newcommand{\ecen}{\end{center}}

\newcommand{\Gev}{\mbox{ GeV}}

\title{Susy-QCD corrections to neutralino pair production in association with a jet}

\author{
Gavin~Cullen\\
Deutsches Elektronen-Synchrotron DESY, Zeuthen, Germany\\
E-mail: \email{gavin.cullen@desy.de}}
\author{
Nicolas~Greiner\\
Max Planck Institut f\"ur Physik, M\"unchen, Germany\\
E-mail: \email{greiner@mpp.mpg.de}}
\author{
Gudrun~Heinrich\\
Max Planck Institut f\"ur Physik, M\"unchen, Germany\\
E-mail: \email{gudrun@mpp.mpg.de}}

\abstract{
We present the NLO Susy-QCD corrections to the production of 
a pair of the lightest neutralinos plus 
one jet at the LHC, 
appearing as a monojet signature in combination with missing energy. 
We fully include all non-resonant diagrams, 
i.e. we do not assume that production and decay
factorise. We derive a parameter point based on the p19MSSM
which is compatible with current experimental bounds and show distributions 
based on missing transverse energy and jet observables.
Our results are produced with the program 
\GOSAM\,\cite{Cullen:2011ac} for automated one-loop calculations in 
combination with MadDipole/MadGraph for the real radiation part.
}
\preprint{MPP-2012-186, DESY 12-239, LPN12-139, SFB/CPP-12-99}

\begin{document}

\section{Introduction}
\label{sec:intro}
With the LHC experiments performing extremely well, we have entered the exciting phase where 
we can investigate the properties of  a new discovery\,\cite{:2012gu,:2012gk}, 
and get exclusion bounds and hopefully also hints 
 for Beyond the Standard Model physics at energy ranges which have never been  probed before.

If the new boson with a mass around 125 GeV is a fundamental scalar, we have to figure out why it is protected from 
higher order corrections of the order of the Planck scale. 
Supersymmetry (for a review see e.g.\,\cite{Martin:1997ns,Peskin:2008nw} and references therein)
offers an elegant explanation for a stabilisation mechanism, 
and in addition contains massive weakly interacting particles which can serve as dark matter candidates.

While the hopes of an early Susy discovery at the LHC 
have withered
as recent LHC measurements have pushed up the lower limits on squark and gluino 
masses considerably\,\cite{ATLAS-CONF-2012-109,:2012rz,:2012mfa,Schumm:1473137,Chatrchyan:2012jx,Kraml:2012er},
there is no stringent lower limit on the mass of the 
lightest neutralino\,\cite{Dreiner:2012ex,Dreiner:2009ic}, 
and the pair production of charginos/neutralinos becomes increasingly important as a ``discovery channel" 
of Supersymmetry.
 
In many Susy scenarios, the neutralino $\tilde{\chi}_1^0$ is assumed to be the 
lightest supersymmetric particle (LSP) and thus is stable if R-parity is conserved. 
Therefore, $\tilde{\chi}_1^0 \tilde{\chi}_1^0$ production, either directly or through 
the decay of heavier neutralinos/charginos, leading to signatures of missing energy 
and energetic jets/leptons, is a process of primary interest in the context of 
current Susy searches.
Hence it is desirable to have predictions which include the NLO Susy-QCD corrections 
to such processes, not only at the level of total cross sections, but also for differential distributions. 
While neutralino pair production without any additional jets is not very illuminating 
from an experimental point of view, the production of neutralino pairs in association with additional jets
is interesting, since the signature ``missing energy + jets" smells like New Physics, and the distributions 
of jet observables can be used  to investigate the nature (e.g. the spin) of the object(s) carrying away the 
missing energy.  

A very clean signal of new physics would be the observation of an excess in 
events involving a very energetic monojet in combination with missing transverse energy. 
Searches for monojets at the LHC at 7 and 8\, TeV have been carried
out already\,\cite{Martinez:2012ie,CMSmonojet,ATLASmonojet,ATLAS-CONF-2012-147}, 
and turned mainly into limits on models involving extra dimensions, assuming that the missing energy is due to 
graviton production. 
If the missing energy is due to neutralinos, studying such processes could provide information 
on the nature and couplings of the LSP and thus on dark matter\,\cite{Allanach:2010pp,Drees:2012dd}.  
Further, monojet signatures are interesting in the context of constraints on  invisible decays   
of the Higgs boson, as the invisibly decaying boson may recoil against 
hard QCD radiation\,\cite{Djouadi:2012zc,Dreiner:2012gx,Englert:2011us}.

Total cross sections for the production of charginos and neutralinos at next-to-leading order in QCD have
been calculated in \cite{Beenakker:1999xh}. Recently,  updates for the LHC at 7 and 8\,TeV and current
popular benchmark points\,\cite{AbdusSalam:2011fc} have been given in \cite{Kramer:2012bx,Fuks:2012qx}. 
Resummation of large logarithms  in the threshold and small-$p_T$ regions has also 
been carried out\,\cite{Debove:2009ia,Debove:2011xj,Debove:2010kf} for the production of gaugino pairs. 
However, predictions for neutralino pair production in association with one or several jets 
in a differential form were, until recently, only available at leading order\,\cite{Allanach:2010pp}. 
The first differential NLO QCD corrections for the final state of missing transverse energy in association with two jets, 
where the missing energy 
stems from the decay of a squark pair into quarks and neutralinos, have been calculated in \cite{Hollik:2012rc}. 
For the case of squark and gluino pair production without decay or additional jets, 
the number of available results beyond the leading order is 
larger than for charginos/neutralinos, because these processes were hoped to be seen 
already at the Tevatron or at early stages of LHC measurements. 
The first NLO calculations\,\cite{Beenakker:1994an,Beenakker:1995fp,Beenakker:1996ch,Bozzi:2005sy}, 
partly entering the code {\tt Prospino}\,\cite{Beenakker:1996ed}, 
were followed by electro-weak corrections\,\cite{Bornhauser:2007bf,Hollik:2007wf,Hollik:2008vm,Germer:2010vn,Germer:2011an},
resummation \,\cite{Beenakker:2009ha,Kulesza:2008jb,Kulesza:2009kq,Beneke:2010da,Kauth:2011vg} 
and NNLO threshold corrections\,\cite{Langenfeld:2009eg,Langenfeld:2012ti}.
NLO QCD corrections to squark and gluino pair production compared to results from LO matrix element plus parton shower
merging have been presented recently in \cite{GoncalvesNetto:2012yt}, based on calculations 
in~\cite{GoncalvesNetto:2012nt,Binoth:2011xi}.

Next-to-leading order predictions involving chargino/neutralino pairs used in experimental analysis were usually obtained by calculating
the NLO K-factors for the total cross sections using e.g. the code {\tt Prospino}\,\cite{Beenakker:1996ed,Beenakker:1996ch}, 
and rescaling the LO predictions accordingly. 
However, it is not at all guaranteed that the same K-factors can be applied globally, in particular in the presence 
of stringent search cuts. 
QCD radiation can change the shape of the distributions and affect the mass and spin measurements 
considerably\,\cite{Horsky:2008yi}. If the supersymmetric spectrum is highly compressed, QCD radiation 
can also seriously affect the exclusion bounds extracted on the basis of 
leading order assumptions\,\cite{Dreiner:2012gx,Dreiner:2012sh}.

In this paper, we calculate the NLO QCD corrections to the production of a pair of the lightest neutralinos in association with one jet.
We include not only resonant contributions from squark decay, but also all  
non-resonant contributions. We present our results in a fully differential form, 
showing distributions for observables involving the jet and missing energy, which can be compared 
straightforwardly to data. The treatment of diagrams involving resonant squarks needs special attention, 
as the NLO real corrections formally also contain diagrams which can be regarded as leading order contributions to a
different process, which is resonant squark pair production and subsequent decay. We also calculate contributions from Higgs production
through a heavy quark or squark loop. However, these contributions are found to be numerically very small.
For our studies we consider the phenomenological MSSM (pMSSM)~\cite{AbdusSalam:2011fc,Djouadi:2002ze,Berger:2008cq}, 
in a variant involving 19 free parameters (p19MSSM). In this framework, we derive a point where 
the mass of the lightest Higgs boson $h$ is $m_{h} = 125.8 \Gev$. The virtual corrections have been calculated 
with the automated one-loop program \GOSAM~\cite{Cullen:2011ac}, where the integrals involving complex masses 
have been called from the integral libraries {\tt Golem95}\,\cite{Binoth:2008uq,Cullen:2011kv} and \texttt{OneLOop}\,\cite{vanHameren:2010cp}. 
The real radiation matrix elements are generated using 
{\tt MadGraph}~\cite{Stelzer:1994ta} and  {\tt MadDipole}~\cite{Frederix:2008hu,Frederix:2010cj}.
 
The paper is organized as follows. 
In Section \ref{sec:calculation}, we describe our method for the calculation of the virtual and real corrections,
in particular the generation of the renormalisation counterterms and the treatment of resonant squarks, 
the latter being further discussed in the Appendix. 
We also include a phenomenological discussion of MSSM parameter points. 
In Section \ref{sec:results} we present our numerical results, before we conclude in 
Section \ref{sec:conclusion}.

\section{Calculational framework}
\label{sec:calculation}

\subsection{Virtual corrections\label{sec:virt}}

The one-loop virtual contribution to the NLO result is calculated using the program
\GOSAM~\cite{Cullen:2011ac}. We use \FEYNRULES ~\cite{Christensen:2008py}
to produce a model file in the \UFO ~\cite{Degrande:2011ua} format that
can be read directly by \GOSAM.

For the virtual amplitude we have $\mathcal{O}(1400)$ diagrams to calculate
for each subprocess. We neglect $b$-quarks in the initial state.
The most complicated diagrams are rank-3 pentagons, with up to 4 internal masses. 
We illustrate two of the pentagon diagrams in 
Fig.\,\ref{fig:virtual}.
We include finite widths in the loop integrals and
therefore we need a basis set of complex integrals, which we call from the libraries 
\GOLEMVC~\cite{Cullen:2011kv} and \texttt{OneLOop}\,\cite{vanHameren:2010cp}.
To calculate the loop amplitude in a numerically robust way, 
\GOSAM \ is able to interchange between different reduction schemes at runtime. 
Our default reduction strategy is to use \SAMURAI~\cite{Mastrolia:2010nb} and, if it fails, 
to reprocess the point with \GOLEMVC~\cite{Cullen:2011kv} using tensorial reconstruction at integrand level~\cite{Heinrich:2010ax}.
Due to the large internal masses and the small squark widths 
present in the integrals, and due to the high rank of the pentagons, 
numerical stability is a nontrivial issue in this process.
Therefore it is crucial that we have this rescue system available during the numerical integration.
We use the dimensional reduction scheme ({\small DRED}) where only the internal momenta are kept in $D$ dimensions. 
We also calculate contributions where  neutral Higgs bosons can be produced
by a loop-induced process. These can then decay to a pair of the lightest neutralinos.

\FIGURE{
\begin{minipage}{30pc}
\includegraphics[width=30pc,angle=0]{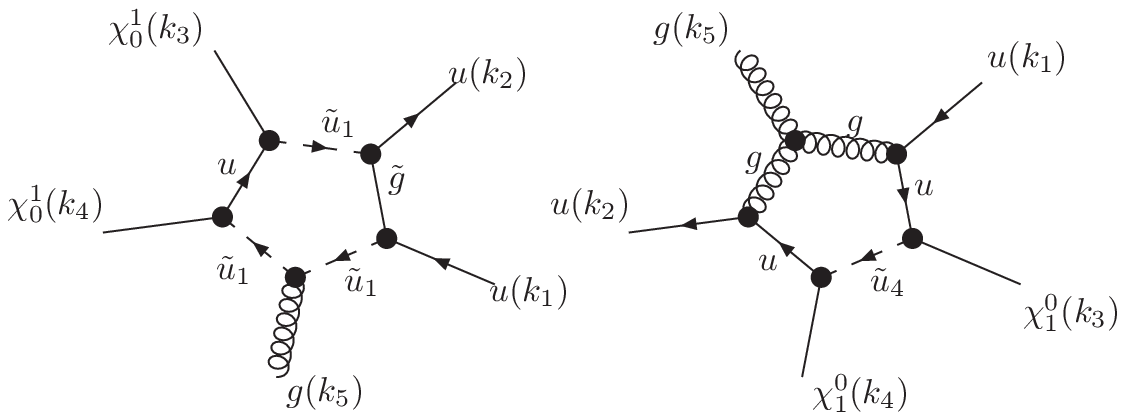}
\end{minipage}
\caption{Two illustrative pentagon diagrams calculated by \GOSAM.\label{fig:virtual}}
}

\subsection{Counter Term Diagrams}

Due to the appearance of self-energy insertions that can have internal masses different
to the mass of the incoming particle we modified the handling of counter term
diagrams from the default \GOSAM \ template file.
In the default \GOSAM \ template file each self-energy insertion in the virtual
diagrams is rewritten as a sum of the original part plus a part that integrates to give the counter term
diagram. Here we pursue a different approach as follows: each diagram containing a self-energy
insertion is ``tagged'' when the diagram topologies are analyzed. This information
is passed through the processing of the diagrams 
and is available when the numerators of the diagrams are reduced. 
The counter term diagram is then reconstructed from the original diagram, by shrinking the self-energy
insertions to a point. This is then reduced alongside the loop diagram and
written to a separate file for numerical evaluation.
The counter term diagrams depend on model dependent renormalisation constants that we calculate
separately. It is planned that this approach
will be incorporated into a future release of \GOSAM. 

\subsection{Real corrections and numerical integration}

The leading order (LO) and the real radiation matrix elements are generated using {\tt MadGraph}
~\cite{Stelzer:1994ta}.
For the subtraction of the infrared singularities we use Catani-Seymour dipoles~\cite{Catani:1996vz},
supple\-men\-ted with a phase space restriction parameter $\alpha$ as proposed in~\cite{Nagy:1998bb}
and implemented in the package {\tt MadDipole}~\cite{Frederix:2008hu,Frederix:2010cj}.
The generation of the various pieces of the code (tree-level, real emission process, subtraction terms)
and their combination with the integration routines was performed in a fully automated way. 
For the numerical integration we used {\tt MadEvent}~\cite{Maltoni:2002qb,Alwall:2007st}, slightly modified
in order to integrate the different NLO contributions.

\subsection{Treatment of diagrams with resonant squarks
\label{sec:doublyresonant}
}

\begin{figure}[htb]
\begin{center}
\includegraphics[width=15pc]{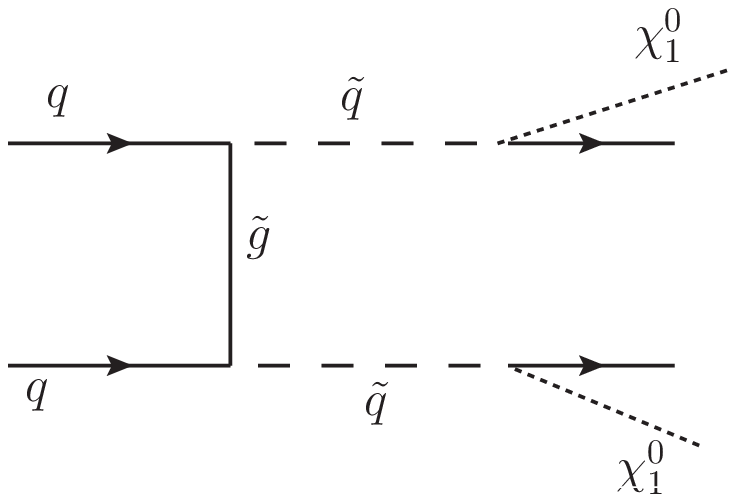} 
\caption{Example of a ``doubly resonant" squark diagram occurring in the NLO real corrections.}
\label{fig:doublyreso}
\end{center}
\end{figure}

For processes involving unstable particles, the proper definition of the set of diagrams 
contributing to the next-to-leading order corrections is not obvious. 
There are problems of double counting as diagrams with additional real radiation 
from the unstable particle 
in the final state can, if it becomes resonant, also be regarded as part of a leading order process 
with the decay already included in the narrow width approximation.
The problem is similar to the case of $W\,t$ and $W^+W^-\,b$ production at NLO, 
where the $W^+\,W^-\,b\bar{b}$ final state occurring in the NLO real corrections, 
if stemming from doubly resonant top decays, can also be viewed as belonging to leading order 
$t\bar{t}$ production and decay.
This problem has been discussed in detail in \cite{Frixione:2008yi,Campbell:2005bb} for the case of 
$W\,t$ production. The case at hand is very similar, with $t\to W\,b$ replaced by 
$\tilde{q}\to \tilde{\chi}_1^0\,q$. Consider for example the leading order diagrams in 
Fig.\,\ref{fig:schannel}. Squark exchange in the t-channel,
as shown in Fig.\,\ref{fig:schannel}(a), cannot lead to any resonance, but in the case of 
s-channel squark exchange shown in Fig.\,\ref{fig:schannel}(b), the squark can become resonant,
and it can be viewed as a diagram for squark production in association with a neutralino, 
with squark decay included in the narrow width approximation.
Now at NLO, when the real radiation of an additional parton is included, 
a new channel opens up, where two squarks can decay resonantly into a quark and a neutralino, 
as shown in Fig.\,\ref{fig:doublyreso}. Close to the resonance, this contribution gets 
quite large, and in fact should rather be counted as a leading order contribution to 
squark pair production with subsequent squark decay, because here we are interested in the radiative corrections 
to the final state of a monojet in association with a neutralino pair. 

\FIGURE{

\begin{minipage}{14pc}
\includegraphics[width=10pc,angle=0]{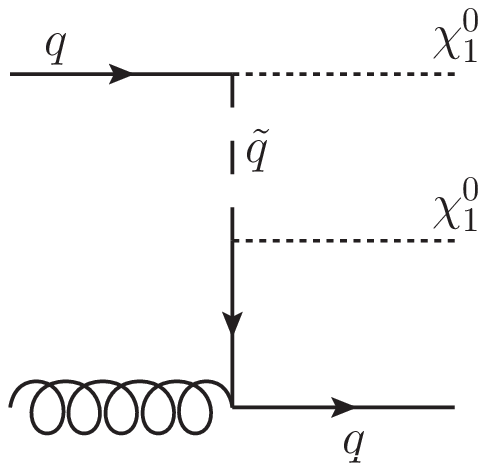}
\end{minipage}\hspace{2pc}%
\begin{minipage}{18pc}
\includegraphics[width=17pc,angle=0]{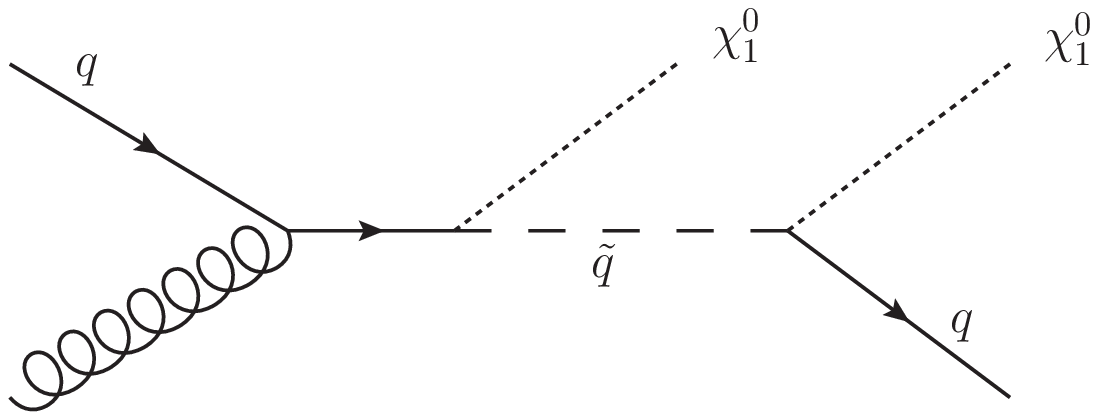}
\caption{Examples of a leading order diagrams with a squark in (a) the t-channel, (b) the s-channel.\label{fig:schannel}}
\end{minipage} 
}




For this reason the calculation was carried out in two different ways. 
In the first approach we take into
account all possible diagrams leading to the required final state consisting of two neutralinos and two
QCD partons. In particular this includes the possibility of having two on-shell squarks.

As it turns out, the real radiation part is vastly dominated by these diagrams, spoiling the
convergence of the perturbation series. The result behaves like a tree-level calculation,
involving one more order  in the strong coupling constant than the tree level for the monojet case. 
The difference to an actual tree-level calculation
is given by the fact that one parton can become unresolved, leading to the signal ``single
jet plus missing energy", while the infrared singularities due to the unresolved parton 
are canceled by the virtual corrections
or absorbed into the parton distribution functions. 
Therefore, even though the calculation which includes these resonant diagrams 
cannot be regarded as a genuine next-to-leading order correction to 
the cross section describing a neutralino pair in association with one jet,
it is still a physically meaningful quantity.
However, from an experimental point of view, 
a complete description of the final state of missing energy plus up to two jets would be 
more useful. This however would require the full NLO calculation of the production 
of a neutralino pair in association with two jets, 
where the jets can either originate from the decay of squarks and gluinos or be produced 
directly from partons in the hard interaction.
This is a very complex task which is beyond the scope of this paper.

In the second approach we follow a strategy proposed in \cite{Frixione:2008yi}, namely we 
remove the diagrams with two squarks in the s-channel from the amplitude. Removing diagrams
from the amplitude generally violates gauge invariance. Ref. \cite{Frixione:2008yi} contains
a study about the impact of violating gauge invariance by such a removal of diagrams, 
where the effects were found  
to be small for commonly used gauges.

For our calculation we assume that the largest contribution
of diagrams with two squarks in the s-channel come from those points in phase space where
both squarks are on-shell, and that off-shell effects are suppressed by a factor of $\Gamma/M$.
Therefore it is sufficient to consider the $2\to2$ process of producing two squarks. 
In this case one
can show that the gauge dependence vanishes for covariant gauges and for a large class of
non-covariant gauges. We give a proof in appendix \ref{sec:appendixA}.

Another solution to the double counting problem based on the subtraction of the resonant diagrams
is the so-called ``prospino scheme"\,\cite{Beenakker:1996ch,Binoth:2011xi},
where in its most recent variant\,\cite{GoncalvesNetto:2012yt} a subtraction term is
introduced  which involves a momentum remapping oriented at
Catani-Seymour mappings to preserve the on-shell conditions for both intermediate and
final state momenta. The subtraction term is then constructed in such a way that it comes into
operation when the intermediate particle goes on-shell.
However, for multi-particle final states where the Born process is already $2\to 3$ or of higher multiplicity, and doubly resonant diagrams appear at NLO, the generalisation of this procedure is not obvious.


\subsection{Phenomenological discussion of Susy parameter points}

The discovery of a boson with mass of about $125 \Gev$~\cite{:2012gu,:2012gk} and a lack of any experimental
evidence for supersymmetry has put many Susy models under strain. 
In particular, the family of the so-called constrained MSSM (cMSSM) 
is increasingly unfavoured in light of the present experimental data~\cite{Buchmueller:2012hv,Strege:2012bt}. 
Therefore we choose a more pragmatic and
experimentally motivated parameterisation of Susy, known as the 
phenomenological MSSM (pMSSM)~\cite{Djouadi:2002ze,Berger:2008cq,AbdusSalam:2011fc}, in a variant involving 19 free parameters (p19MSSM). 
In the pMSSM, no underlying Susy breaking scenario is chosen, and gauge coupling unification is not postulated.
However, it is assumed that there are no sources of CP violation and generation
mixing beyond the Standard Model ones, so that all off-diagonal elements in the sfermion
mass matrices are equal to zero, and  the first and second-generation soft terms are equal.
We  will further assume that  the LSP is the lightest neutralino.

In such models we can corner the parameter space where the squarks, gluino and neutralinos have masses close together. 
This region is of particular
interest for the process considered here, as the squark masses can be low enough for 
the signal to be 
significant at LHC energies while evading the current LHC exclusion regions.

Compressed spectra of this type were also studied
in~\cite{Dreiner:2012gx,Dreiner:2012sh,Dreiner:2012ex} 
where the
search for Susy through the recoil of light Susy particles against initial state radiation
is discussed.
Our choice of Susy parameter point can be viewed as qualitatively like the ``Equal
Mass'' scenario in Ref. \cite{Dreiner:2012sh}. In this situation we expect 
to see soft jets from the decays of the squarks to the neutralino.

For our results we choose a modification of the parameter point p19MSSM1.1
which we call the p19MSSM1Amod.
The p19MSSM1 line was introduced in~\cite{AbdusSalam:2011fc} and was constructed as a benchmark for 
these compressed Susy scenarios, and is labelled by an integer $N$, i.e. p19MSSM1.$N$. 
The p19MSSM1 line is controlled by one parameter: the gaugino mass parameter 
$M_{1}$.
The two lightest generation sfermion masses 
$M_{\tilde{f}_{1,2}}$ and the mass
of the gluino, $M_{3}$, are fixed at $1.2 M_{1}$. p19MSSM1.$N$ is defined for $N=1$ at the value
$M_{1} = 300 \Gev$ and this is increased by $100 \Gev$ for each subsequent value of $N$.
We effectively decouple all other particles in the model
by setting the other mass parameters to a higher scale, here $2500 \Gev$. We choose a value of 10 for $\tan \beta$.

\begin{figure}[htb]
\begin{center}
\includegraphics[width=20pc]{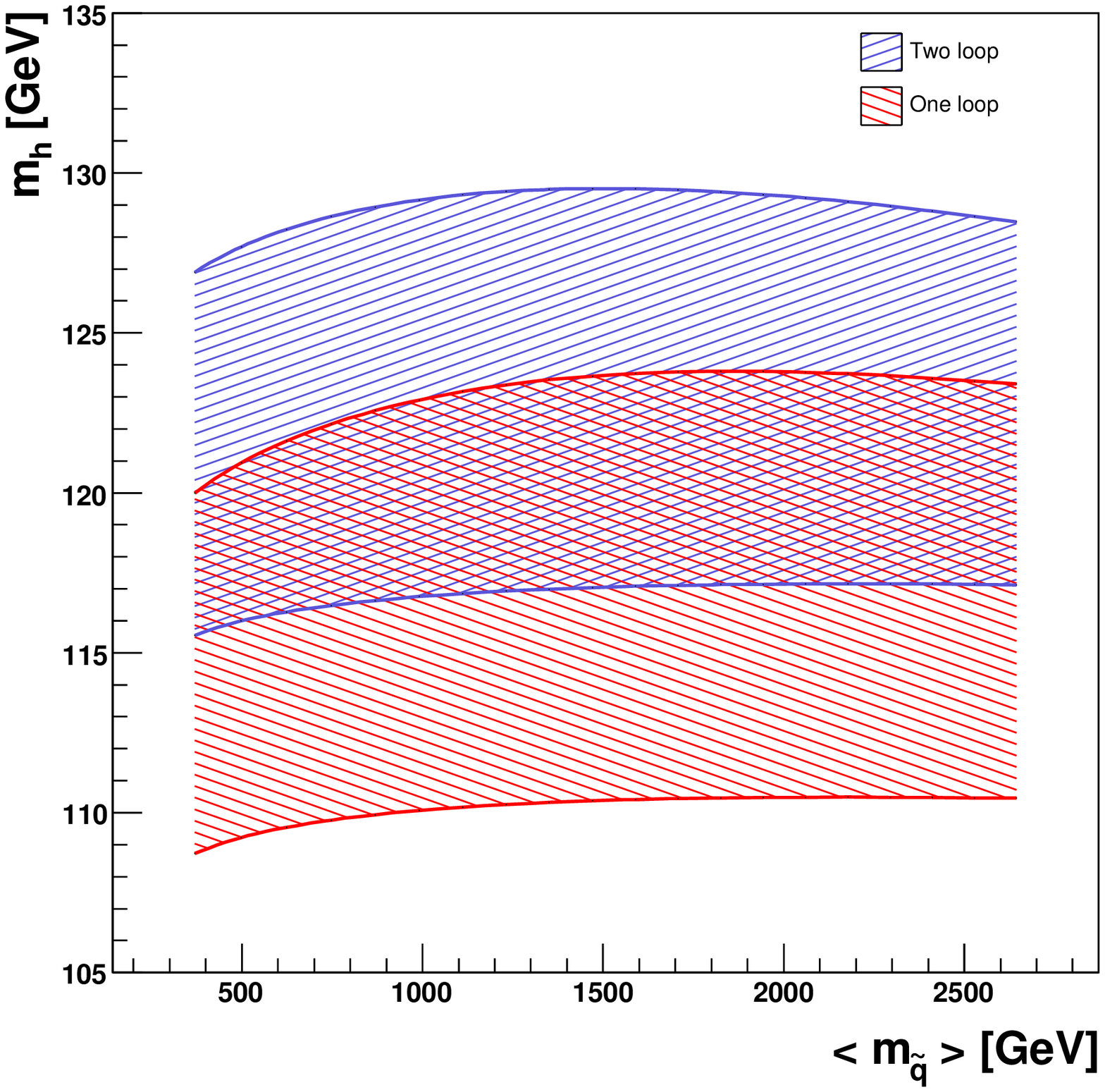} 
\caption{The lightest Higgs boson in the MSSM depends on the parameter $A_{t}$ through
one-loop and two-loop corrections. Here, the band illustrates how $m_{h}$ varies with $A_{t}$ in the 
benchmark scenario p19MSSM1. This variation is examined over a range of input parameters $M_{1}$ which
changes the spectrum of the light squark masses. The x-axis is the average value of the light squark 
masses.}
\label{fig:varyAt}
\end{center}
\end{figure}

Our choice modifies p19MSSM1.1 in two ways. Firstly, we make contact with the point p19MSSM1A,
given in~\cite{Hollik:2012rc}, by setting the heavier Higgs bosons mass inputs, 
$\mu$ and $m_{A}$, to the higher scale. Like the heavier squarks these particles are
effectively decoupled.
Secondly, we further modify this point by choosing $A_{t} = 5000$ such that the mass of the lightest
Higgs $h$ in our model can be identified with the boson observed at the LHC with mass
$m_{h} = 125.8 \Gev$. 
We illustrate the effect that varying $A_{t}$ has on the mass of the lightest Higgs boson from 
$A_t=0$ to the maximum Higgs mass in Fig.\,\ref{fig:varyAt} over a range of the gaugino mass parameter $M_{1}$.
We do not plot the theory uncertainty coming from unknown higher order
corrections nor the uncertainty from the input parameters. The Susy masses that we use for our calculation, at 
this point in parameter space, are given explicitly in Table \ref{SUSYparameters}.

The particle spectrum was produced using \SOFTSUSY~\cite{Allanach:2001kg} and then the decay
widths calculated using the package \SUSYHIT~\cite{Djouadi:2006bz} in which we used the packages
HDECAY~\cite{Djouadi:1997yw} and \SDECAY~\cite{Muhlleitner:2003vg}. \GOSAM \
can read input cards provided in the \SLHA \ format~\cite{Skands:2003cj,Allanach:2008qq} 
allowing one to change the Susy parameter point without recompiling the code.

\subsection{Higgs Contribution to the Signal}

As mentioned in Section \ref{sec:virt} we also calculate the contribution arising from Higgs production through 
heavy quark and squark loops with subsequent Higgs decay to a neutralino pair.
We illustrate these types of
diagrams in Fig.\,\ref{fig:Higgscontrib}.
To quantify their contribution to the total cross section
we can easily isolate these types of diagrams using the diagram filtering system in \GOSAM. We find 
the Higgs boson contribution to the total cross section to be negligible, so these diagrams
are not included in the results shown in Section \ref{sec:results}.

\FIGURE{
\begin{minipage}{12pc}
\includegraphics[width=17pc,angle=0]{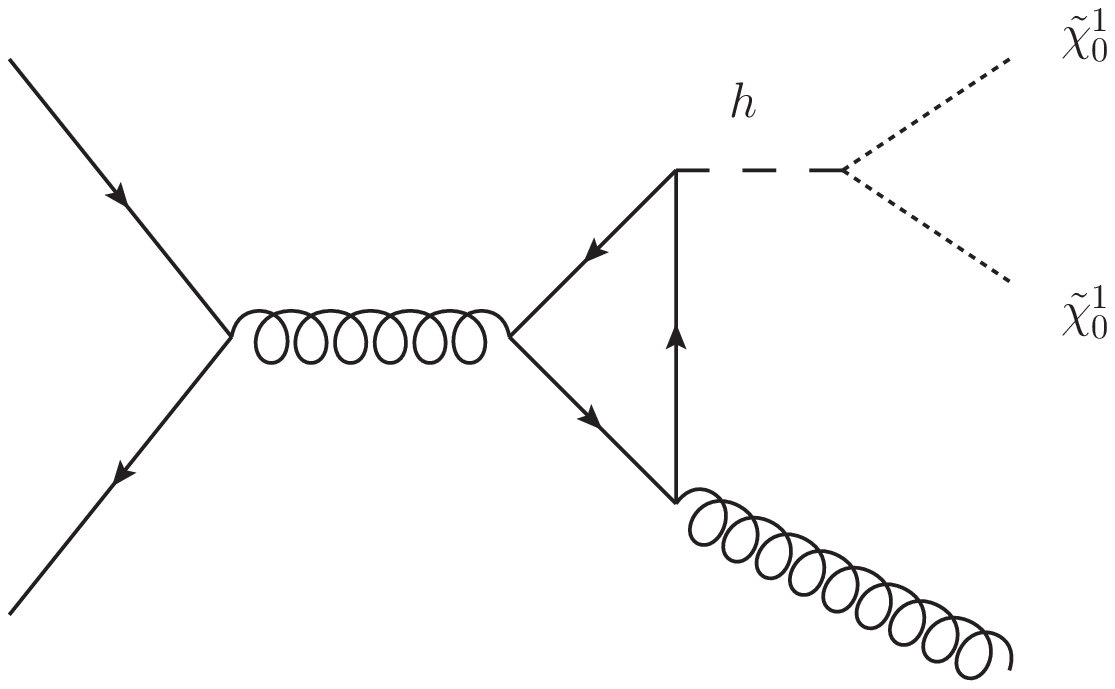}
\end{minipage}\hspace{5pc}%
\begin{minipage}{12pc}
\includegraphics[width=17pc,angle=0]{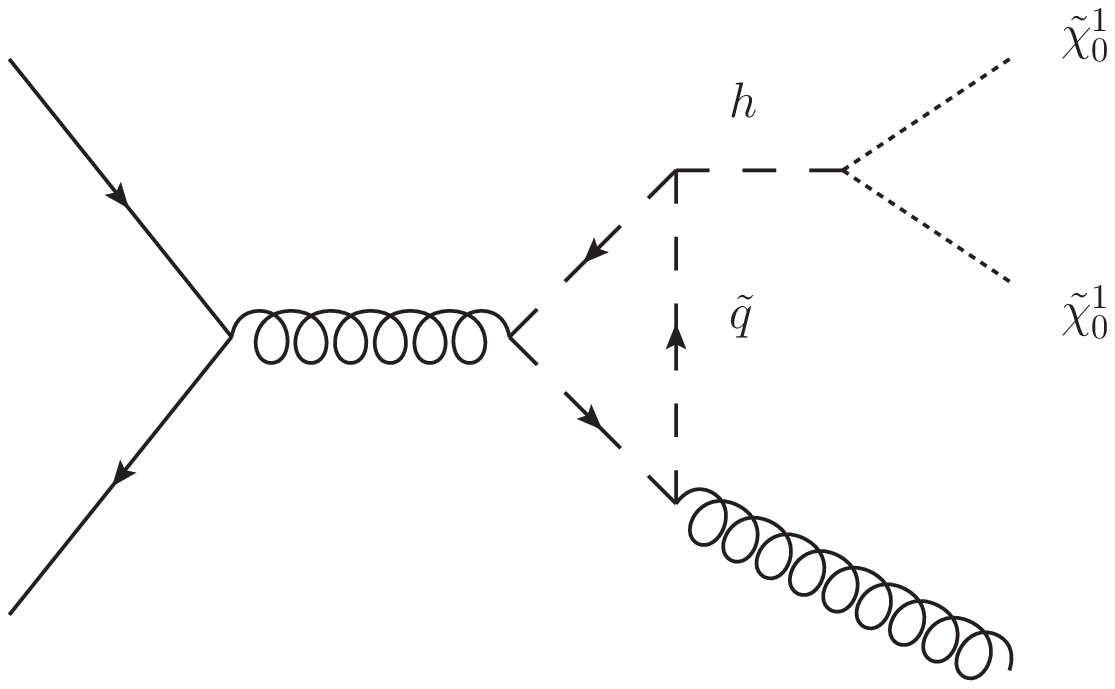}
\end{minipage}
\caption{Examples of a loop diagrams involving the MSSM Higgs bosons. Here $h$ stands for any of the
MSSM Higgs bosons in the set $\{h, H, A\}$.\label{fig:Higgscontrib}}
}

\subsection{Checks on the Result}

We have checked that after UV renormalisation, all poles from the virtual contributions cancel 
with the poles from the infrared insertion operator~\cite{Catani:1996vz} in the real radiation.
We have also checked the unrenormalised virtual matrix element against the program
\FEYNARTS / \FORMCALC ~\cite{Hahn:1998yk,Hahn:2000kx,Hahn:2001rv}. We found agreement
for the partonic subprocess $u g \rightarrow \chi_{1}^{0} \chi_{1}^{0} u$. 
All other subprocesses can be found by exploiting crossing symmetry.

\FIGURE{
\begin{minipage}{10cm}
\includegraphics[width=19pc,angle=0]{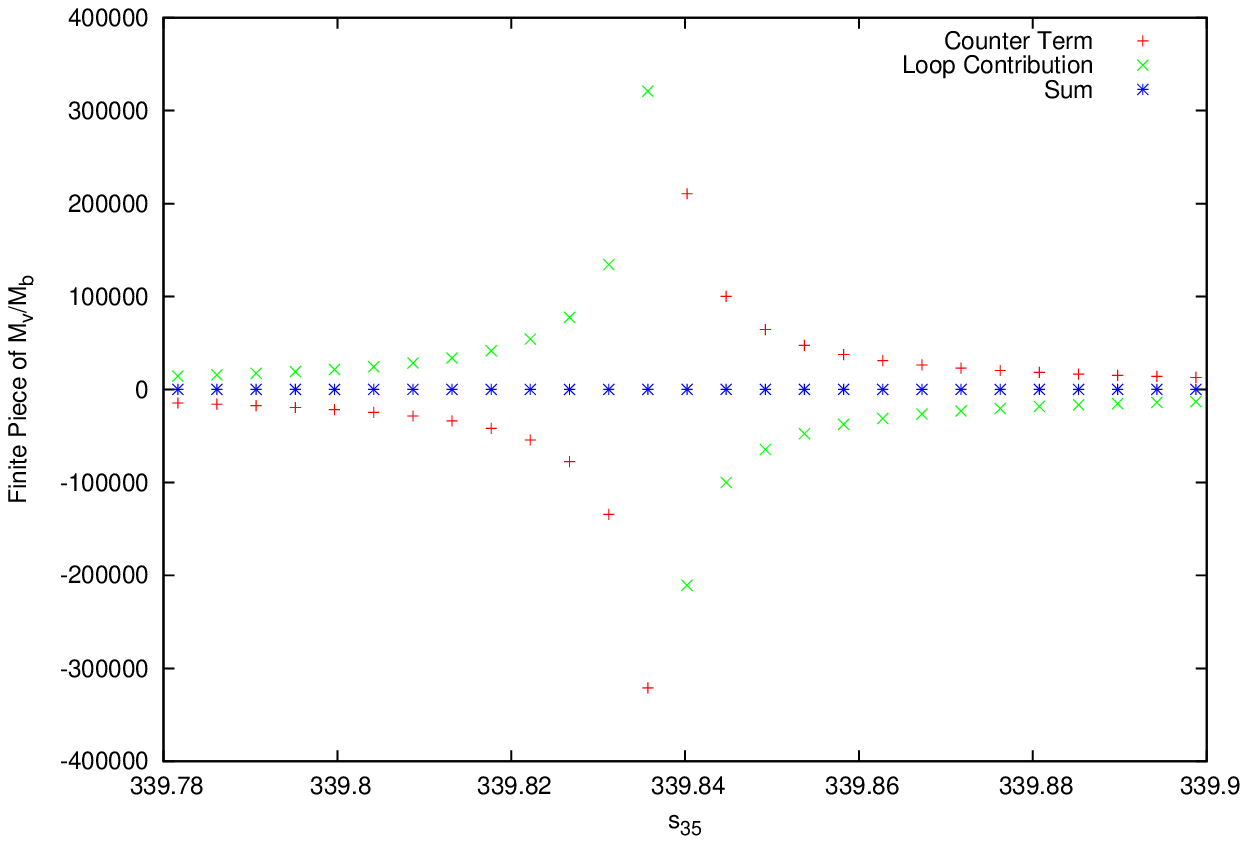}
\caption{Cancellation of the finite remainders of the 
UV divergent parts of the amplitude with the corresponding counter terms in the resonance region. \label{fig:counterterms}}
\end{minipage}
}

Furthermore, we checked our renormalisation procedure by investigating the cancellation 
of the finite remainders of the UV divergent loop contributions with the counter terms, 
as illustrated in Fig.~\ref{fig:counterterms}.

\newpage

\section{Phenomenological results}
\label{sec:results}

In this section we present a selection of phenomenological results 
for proton proton collisions at the LHC at $8$\,TeV.

\subsection{Setup and cuts}
For all the results and distributions shown in this section we have used the
parameters listed in Tables \ref{parameters} and \ref{SUSYparameters}.
\begin{table*}[htb] 
\begin{center} 
{\small
\begin{tabular}{|l|l|}
\hline
\multicolumn{2}{|c|}{Standard Model Parameters}\\
\hline
$M_Z = 91.188\Gev$ & $\Gamma_Z = 2.4952\Gev$ \\
$M_t = 173.4\Gev$ & $\Gamma_t = 1.4384\Gev$ \\
$\alpha(M_Z) = 1/127.934$   & \\
\hline
\end{tabular} 
 }
\end{center}
\caption{Standard Model parameters used for the phenomenological study.
\label{parameters}
}
\end{table*}

\begin{table*}[htb]

\begin{center} 
{\small
\begin{tabular}{|l|l|}
\hline
\multicolumn{2}{|c|}{SUSY Parameters}\\
\hline
$M_{\tilde{\chi}_1^0} = 299.5$ & $\Gamma_ {\tilde{\chi}_1^0}= 0 $ \\ 
$M_{\tilde{g}}    = 415.9$ & $\Gamma_ {\tilde{g}}= 4.801$ \\ 
$M_{\tilde{u}_L} = 339.8$ & $\Gamma_ {\tilde{u}_L}= 0.002562$ \\ 
$M_{\tilde{u}_R} = 396.1$ & $\Gamma_ {\tilde{u}_R}= 0.1696$ \\ 
$M_{\tilde{d}_L} = 348.3$ & $\Gamma_ {\tilde{d}_L}= 0.003556$ \\ 
$M_{\tilde{d}_R} = 392.5$ & $\Gamma_ {\tilde{d}_R}= 0.04004$ \\ 
$M_{\tilde{b}_L} = 2518.0$ & $\Gamma_ {\tilde{b}_L}= 158.1$ \\ 
$M_{\tilde{b}_R} = 2541.8$ & $\Gamma_ {\tilde{b}_R}= 161.0$ \\ 
$M_{\tilde{t}_L} = 2403.7$ & $\Gamma_ {\tilde{t}_L}= 148.5$ \\ 
$M_{\tilde{t}_R} = 2668.6$ & $\Gamma_ {\tilde{t}_R}= 182.9$   \\
\hline
\end{tabular} 
 }
\end{center}
\caption{Masses and widths of the supersymmetric particles for the benchmark point used. The second
generation of squarks is degenerate with the first generation of squarks. All parameters are given in$\Gev$.\label{SUSYparameters}
}
\end{table*}

The weak mixing angle is calculated from the $W$ and $Z$ masses.
The strong coupling constant and its running are determined by the set of parton distribution functions. 
We used an NLO pdf set from  NNPDF2.3 \cite{Ball:2012cx}, where the values for $\alpha_s$ at leading order 
and next-to-leading order are given by
$$
\alpha_{s}(M_Z) = 0.119\;, 
$$
and the running is calculated at one loop for the tree-level result and at two loops for the next-to-leading order
 parts. As we neglect initial state $b$-quarks, we use the $N_f=4$ version of the pdf set.
 Further, we assume flavour-diagonal Susy-QCD couplings.

For the  jet clustering we used an anti-$k_T$ algorithm~\cite{Cacciari:2008gp} with a cone size
of $R=0.4$ provided by 
the {\tt FastJet} package \cite{Cacciari:2011ma,Cacciari:2005hq}.
We choose $\mu=H_T/2$ for our central scale, where we define $H_T$ as $H_T=\sum_i E_{T,i}$ with $i$ running over the momenta of the two neutralinos and the jet(s).

We  use  the following set of cuts  
\beq
p_{T}(\rm{leading\, jet}) \ge 100 \Gev, \quad |\eta_j| \le 4.5\;. 
\eeq
In addition we impose a cut on the missing transverse energy of
\beq
E_{T,\rm{miss}} \ge 85\Gev.
\eeq
We also impose a jet veto of 30\Gev{} on a second jet which at NLO originates from the $2\to 4$ part
of the real radiation corrections, as discussed in Section \ref{sec:doublyresonant}.
Our relatively low cut on the transverse missing energy is motivated by the fact that the neutralinos $\tilde{\chi}_1^0$  could be rather light, 
and therefore the requirement of very large $E_{T,\rm{miss}}$, which is well motivated in searches for graviton production 
in association with  monojets, could be too restrictive in the case of neutralino pair production in association with one jet.
In this case the neutralinos do not originate from long cascades of heavier objects with additional missing energy produced along the cascade.



\subsection{Numerical results}
\label{sec:dist}

In this section we show distributions for the observables $p_T^{j}$, the transverse momentum of the jet, 
the missing transverse energy $E_T^{\rm{miss}}$, and the angle $\phi(\vec{p}^{\,\rm{miss}},\vec{p}^j)$,
where $\vec{p}^{\,\rm{miss}}$ is defined as minus the vector sum of the visible particles in the event,
and $\vec{p}^j$ is the momentum of the leading jet.
We show two types of distributions for each observable: one where each distribution is normalized to one
in order to exhibit the difference in shape, and another with absolute values.
For the normalized histograms in Figs. \ref{fig:ptc}, \ref{fig:ETmissc} and \ref{fig:deltaphic}, 
we show results for both approaches, 
the one including the doubly resonant diagrams and the one with diagram removal. 

Investigating the behaviour of the cross sections under scale variations, we observe the following.
As expected, the results including the doubly resonant diagrams show no improvement of the scale uncertainty at NLO, 
because they are completely dominated by the $2\to 4$ real radiation, and therefore the scale dependence is not compensated
by the virtual contributions.
The  case where these diagrams are removed is still dominated by the 
new channels opening up in the NLO real radiation contributions. 
Therefore, in this case  we do not find a stabilisation of the scale dependence either.
However, the cross sections are sizeable.
Using $\mu=\mu_R=\mu_F$ and varying between $ H_T/4\leq \mu\leq 2\,H_T$, we find 
20 - 30 fb  for the LO cross section, while the NLO subtracted cross section 
amounts to about 100 fb  for the central scale, and the one including doubly resonant diagrams
to 960 fb  for the central scale. This means that the point p19MSSM1Amod considered here 
could in principle be tested with the data accumulated so far.

For the results including the doubly resonant diagrams, it is pointless to determine a K-factor, as 
in this case it is not well defined to which leading order process the higher order terms should be attributed, 
as explained in Section \ref{sec:doublyresonant}. 



\FIGURE{
\begin{minipage}{10cm}
\includegraphics[width=11.cm]{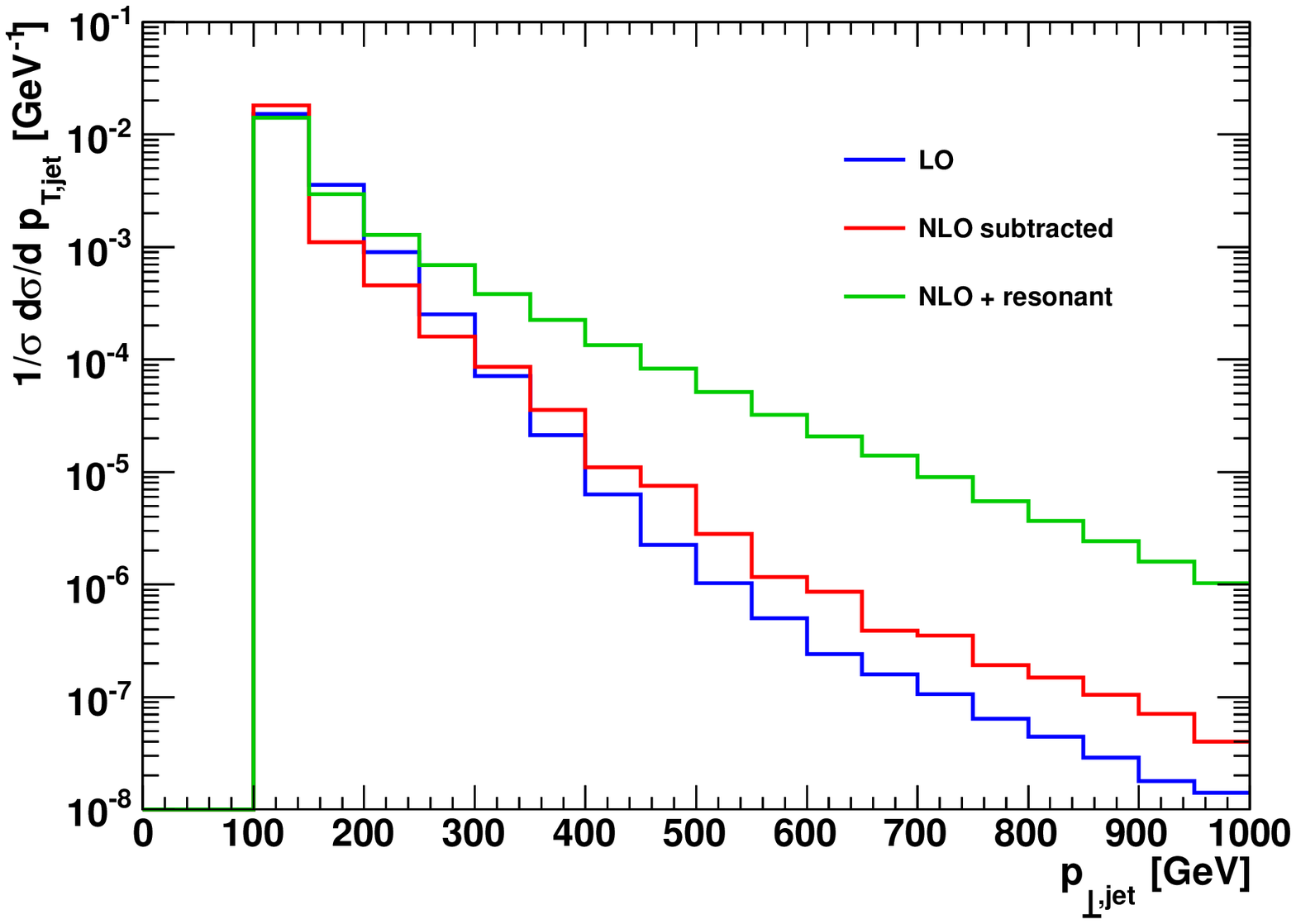} 
 \caption{Normalized distributions showing the transverse momentum distribution of the leading jet
for the process $pp\to \tilde{\chi}_1^0\tilde{\chi}_1^0$+jet at $\sqrt{s}=8$\,TeV, comparing the cases where the resonant diagrams are included to the ones where they are subtracted.
\label{fig:ptc} }
\end{minipage}
}

 \FIGURE{
\begin{minipage}{10cm}
\includegraphics[width=10.cm]{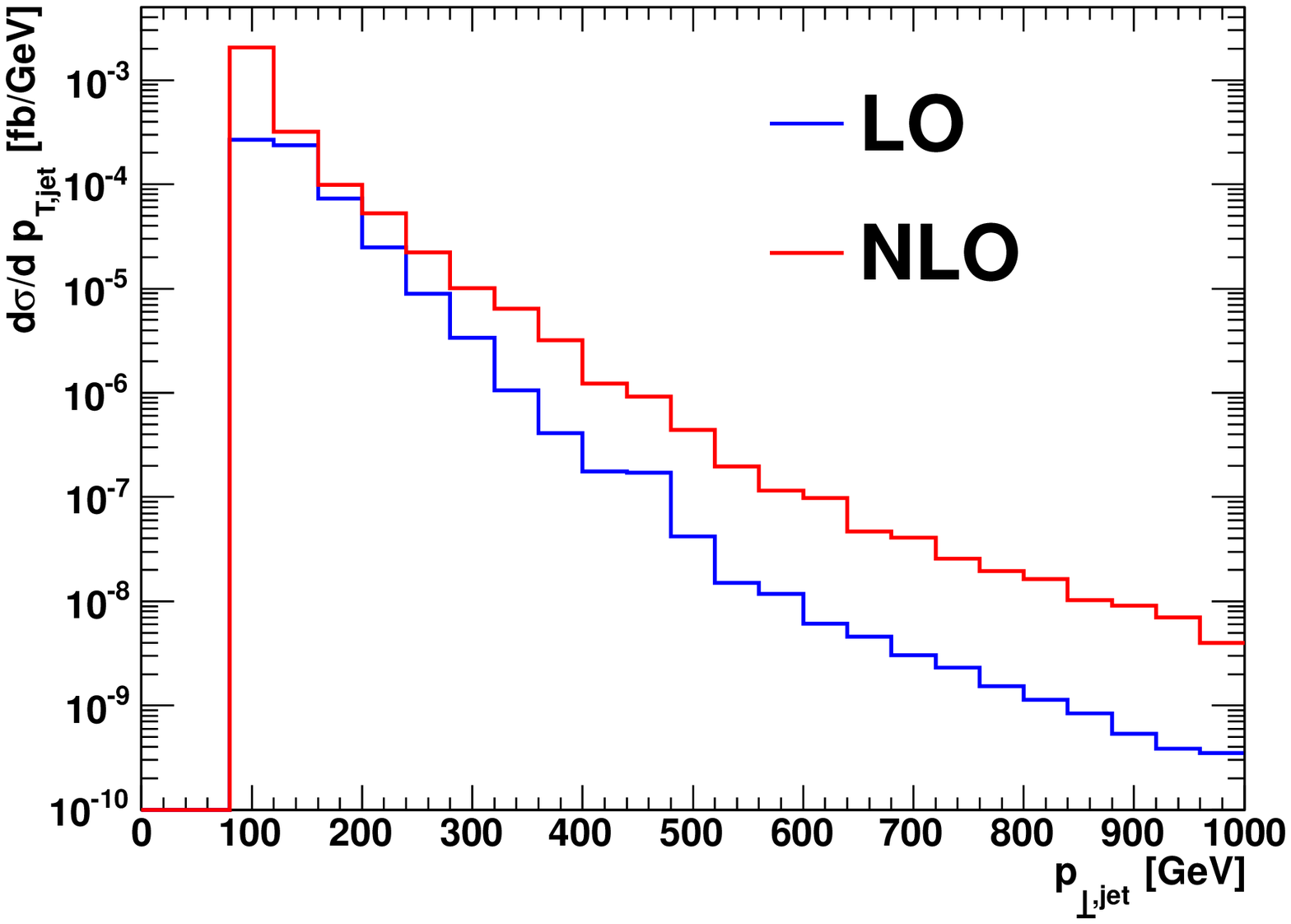} 
 \caption{Transverse momentum distribution of the leading jet
for the process $pp\to \tilde{\chi}_1^0\tilde{\chi}_1^0$+jet at $\sqrt{s}=8$\,TeV.\label{fig:pt} }
\end{minipage}}

\FIGURE{
\begin{minipage}{10cm}
\includegraphics[width=11.cm]{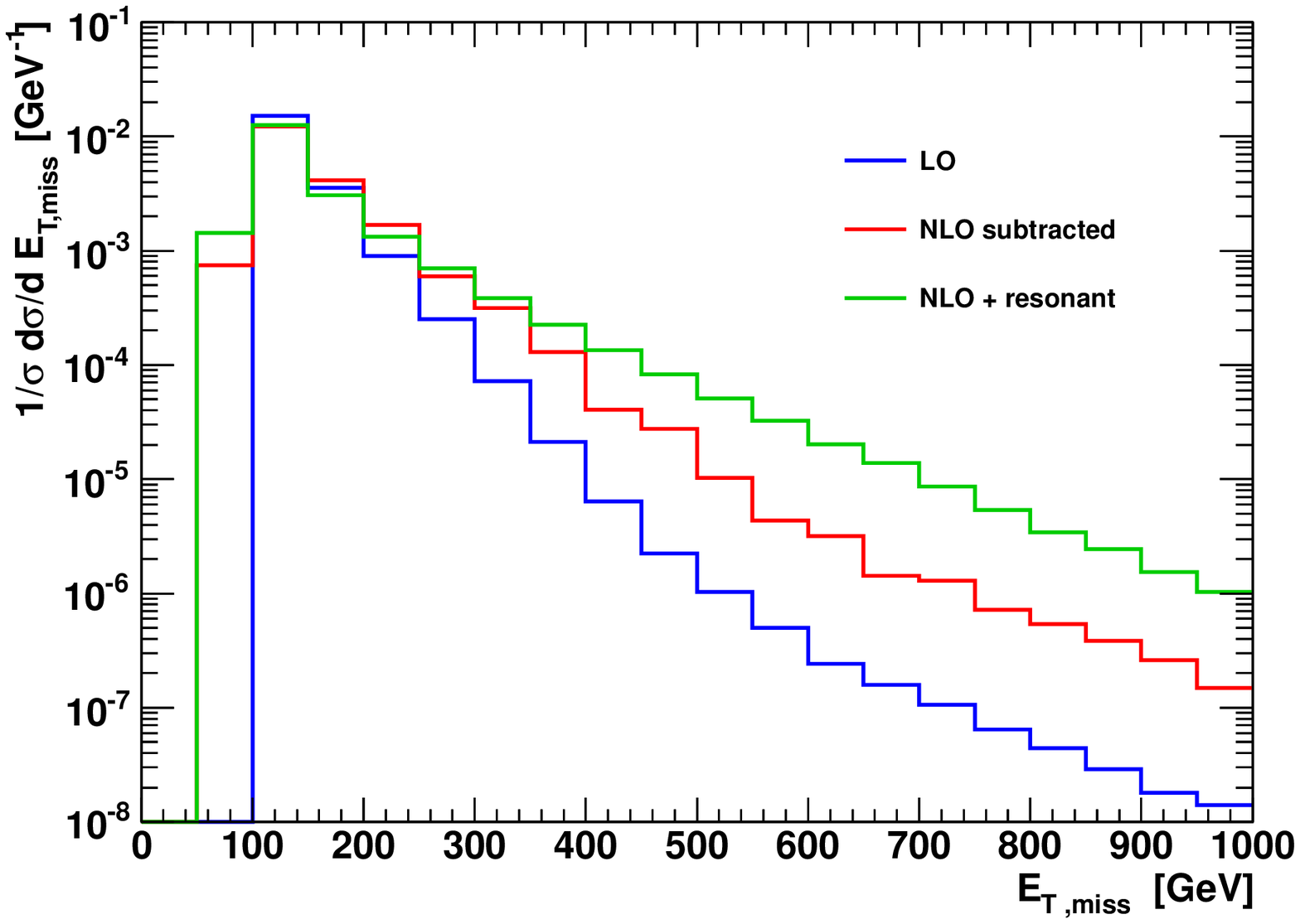} 
 \caption{Normalized distributions showing the  missing transverse energy $E_T^{\rm{miss}}$,
 comparing the cases where the resonant diagrams are included to the ones where they are subtracted.}
\label{fig:ETmissc}
\end{minipage}
}

\FIGURE{
\begin{minipage}{10cm}
\includegraphics[width=10.cm]{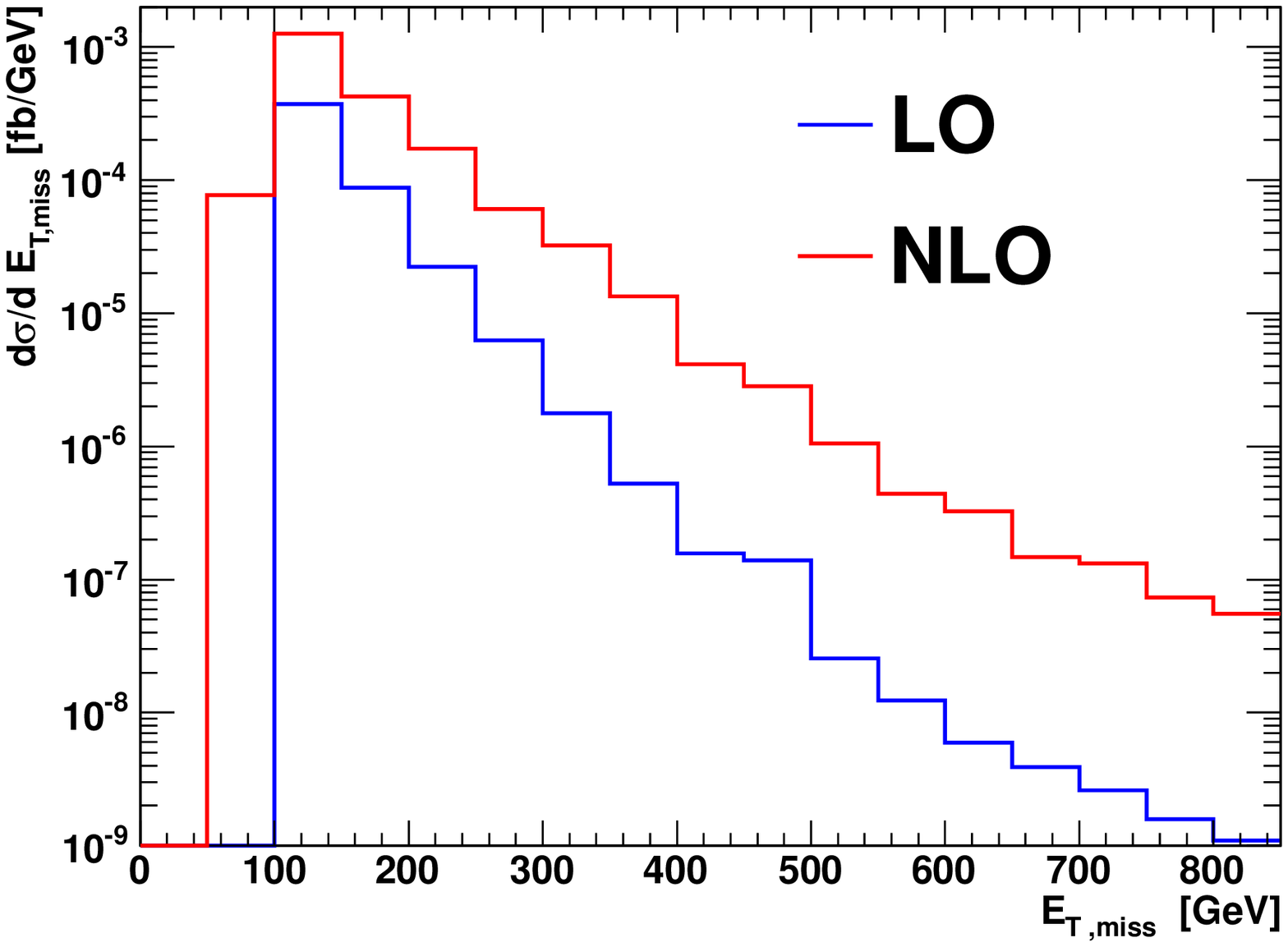} 
 \caption{Distribution showing the  missing transverse energy $E_T^{\rm{miss}}$ for the process $pp\to \tilde{\chi}_1^0\tilde{\chi}_1^0$+jet at $\sqrt{s}=8$\,TeV.}
\label{fig:ETmiss}
\end{minipage}
}

\FIGURE{
\begin{minipage}{10cm}
\includegraphics[width=11.cm]{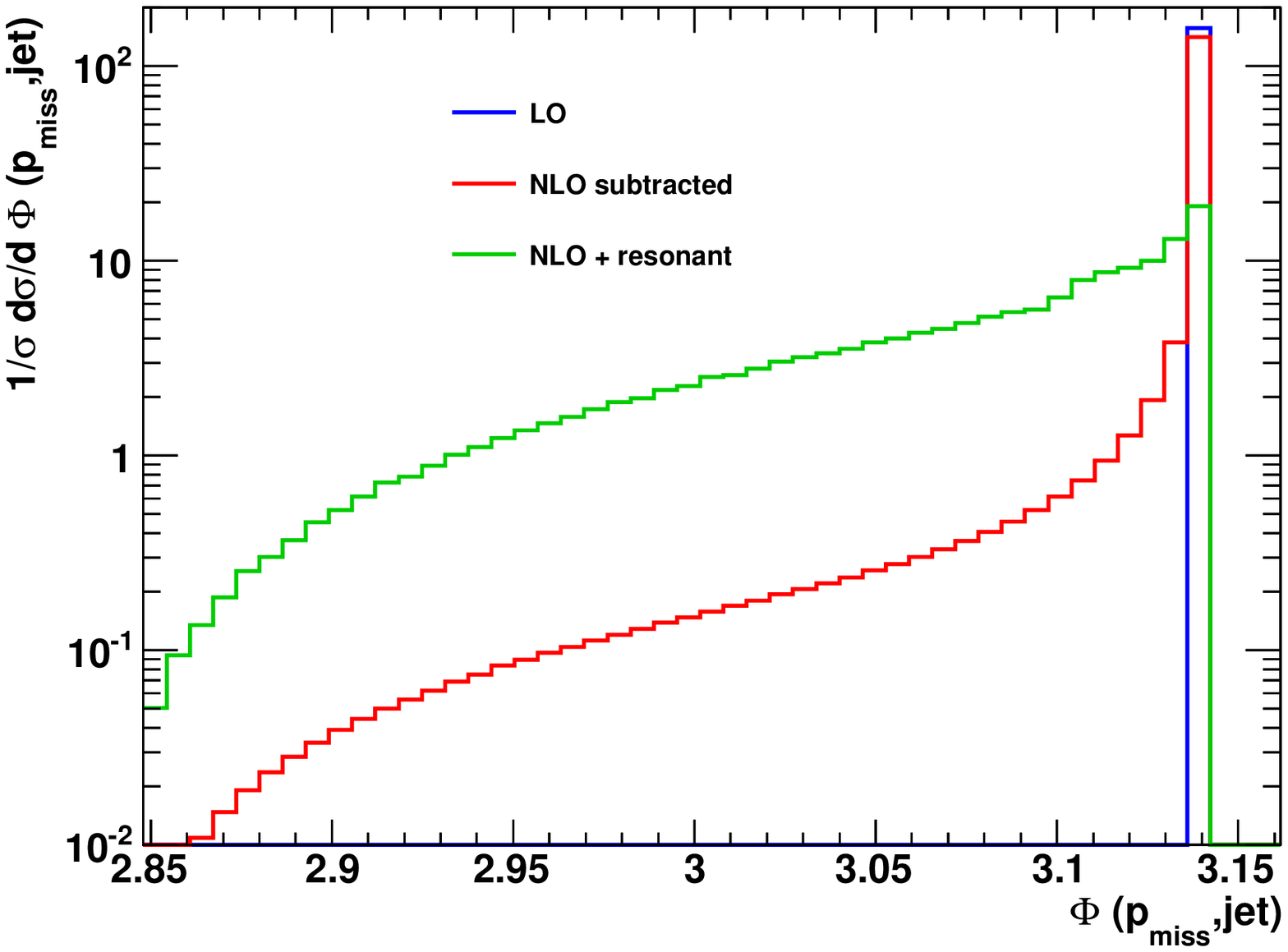} 
\caption{Normalized distributions showing the angle $\phi(\vec{p}^{\rm{miss}},\vec{p}^j)$, 
comparing the cases where the resonant diagrams are included to the ones where they are subtracted.}
\label{fig:deltaphic}
\end{minipage}
}

\FIGURE{
\begin{minipage}{10cm}
\includegraphics[width=10.cm]{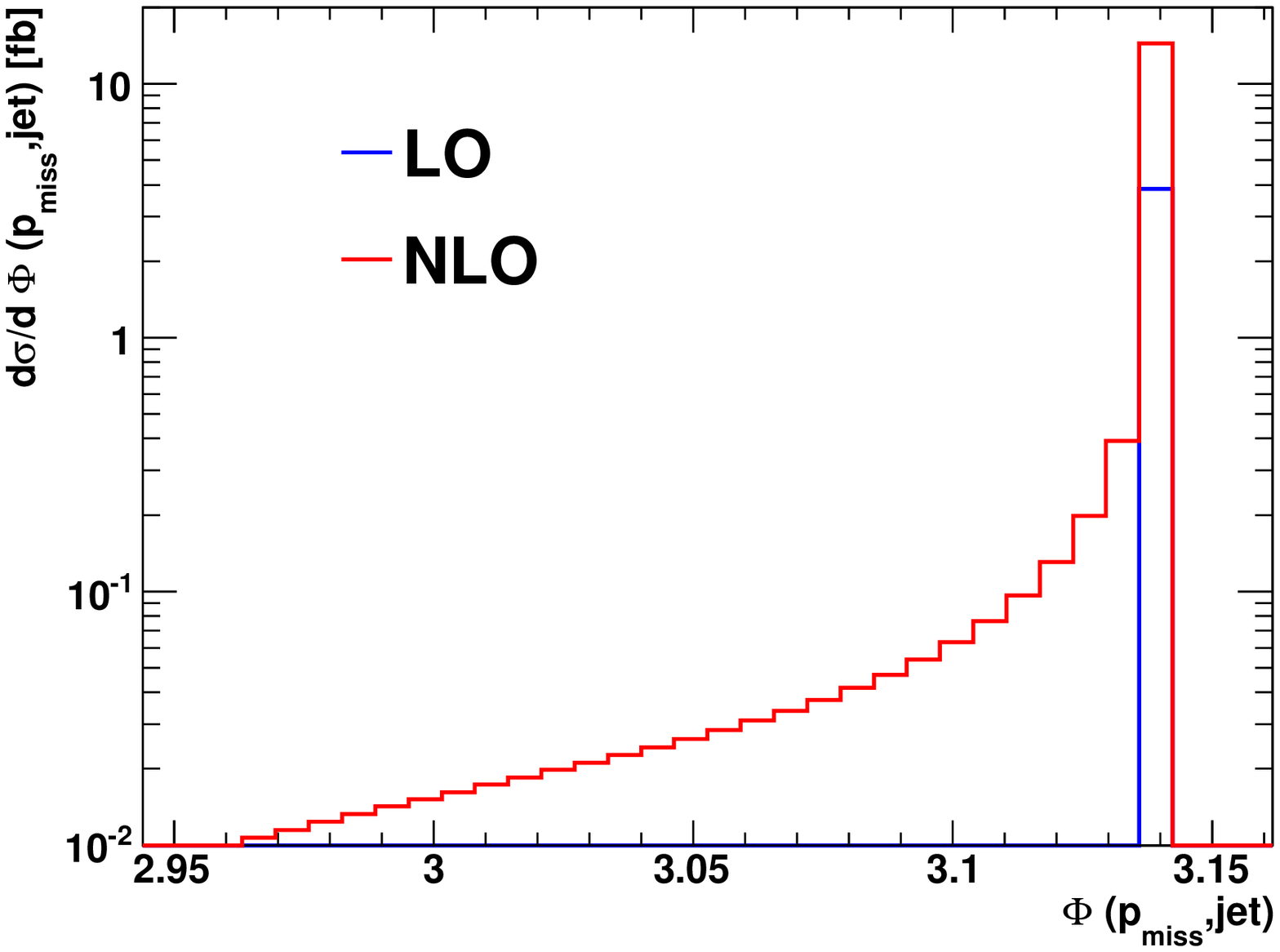} 
\caption{The angle $\phi(\vec{p}^{\rm{miss}},\vec{p}^j)$ between the leading jet and missing momentum  vectors, 
where the latter is defined as the negation of the visible momenta in the event.}
\label{fig:deltaphi}
\end{minipage}
}
For the subtracted case, where we show absolute as well as normalized results, the K-factors are still quite large, as can be seen from 
Figs. \ref{fig:pt}, \ref{fig:ETmiss} and \ref{fig:deltaphi}. 
This can be understood as being mainly due to new channels opening up in the NLO real radiation part,
in the presence of an additional QCD parton.
The distributions also show that the K-factors are  not uniform, 
which implies that the ``standard" procedure to use K-factors based on total cross sections calculated at NLO 
can be misleading. For the angle $\phi(\vec{p}^{\,\rm{miss}},\vec{p}^j)$ between the  jet and the missing momentum,
shown in Figs.~\ref{fig:deltaphic} and \ref{fig:deltaphi}, NLO is obviously the first order 
where the fixed order prediction makes sense, because at LO the vector sum of the neutralino momenta will  always 
be back-to-back to the recoiling jet.

The large K-factors can partly be attributed to the fact that at NLO, a considerable number of new partonic channels 
are opening up: the gluon-gluon initiated processes, and the ones initiated by $uu,dd,\bar{u}d,u\bar{d}$ 
are all absent at leading order (where only $q\bar{q}$ and $qg/\bar{q}g$ are present)
and the new channels together make up almost 50\% of the cross section. 
If we roughly estimate the K-factor which would result 
from partonic initial states which are already present at LO, 
it would amount to $K\sim 2.3$.  
Further, we investigated the point SPS1a~\cite{Allanach:2002nj} just for reference, and found that for this point the K-factors 
are also smaller. This can be attributed to the fact that for the compressed spectrum we are considering here, 
the widths of the first generation squarks are very small, leading to larger contributions if the squarks are
close to being on-shell.



\section{Conclusions}
\label{sec:conclusion}
We have calculated the NLO Susy-QCD corrections to the production of a pair of 
the lightest neutralinos in association with one jet.
We did not use the approximation of factorising production and decay, but 
fully included all non-resonant contributions. Contributions from Higgs production
through a heavy quark or squark loop were calculated and found to be numerically negligible.
The calculation has been performed using two different approaches to treat the doubly resonant 
diagrams appearing in the NLO real radiation contribution: one is based on diagram removal
and the other based solely on a veto on the second jet. In the latter approach, the K-factors 
are obviously very large as the whole result is dominated by a reaction which can 
also can be viewed as the Born level for a different process (resonant squark pair production with a 
subsequent factorisable decay into a neutralino and a jet). 
We present our results in a fully differential form, based on the experimentally accessible 
jet and missing $E_T$ observables.

On a technical level, to the best of our knowledge, this is the first $2\to 3$ NLO calculation 
within the MSSM which includes full off-shell effects, as well as complex masses. 
Even though we only consider the
phenomenological MSSM (p19MSSM) here, we would like to emphasize that our setup is largely  automated, 
using the public one-loop program 
\GOSAM{} in combination with {\tt MadGraph}, {\tt MadDipole} and {\tt FeynRules}, 
such that other parameter points, and even other models Beyond the Standard Model, 
can be studied as well within the same framework.

\section*{Acknowledgements}
We would like to thank Wolfgang Hollik, Jonas Lindert, Edoardo Mirabella, Davide Pagani 
and the members of the GoSam collaboration for various useful discussions.
We also acknowledge use of the computing resources at the Rechenzentrum Garching.
The work of G.C. was supported by DFG
Sonderforschungsbereich Transregio 9, Computergest\"utzte Theoretische Teilchenphysik.
We also acknowledge the support of the Research Executive Agency (REA)
of the European Union under the Grant Agreement number
PITN-GA-2010-264564 (LHCPhenoNet).

\renewcommand \thesection{\Alph{section}}
\renewcommand{\theequation}{\Alph{section}.\arabic{equation}}
\setcounter{section}{0}
\setcounter{equation}{0}

\appendix
\section{Gauge dependence}
\label{sec:appendixA}
In this appendix we examine the gauge dependence of the diagrams that have been removed from the amplitude in
the real emission part as discussed in \ref{sec:doublyresonant}. We show that this gauge dependence vanishes 
for covariant gauges and for a large class of non-covariant gauges.

The only diagrams, once omitted, that can lead to a dependence on the choice of gauge are of the type shown in Fig.\ref{squark}.
In this diagram there is an s-channel gluon which decays into a squark-antisquark pair. As the biggest contribution to the cross section
comes from the parts of the phase space where the two squarks are on-shell, it is sufficient for our argument 
to consider the 2$\to$2 proccess
of squark pair production and neglect the subsequent decay of the squarks.
We denote the incoming momenta of the quarks as $q_1,q_2$ and the outgoing momenta of the squarks as $p_1,p_2$.
In the following we neglect overall prefactors like color factors and coupling constants as they are irrelevant for our
argument. The same holds for factors of $i$ and any minus signs.
The amplitude of the 2$\to$2 process can be written as
\begin{equation}
{\cal{M}}\sim \bar{v}(q_1)\; \gamma^{\mu}\; D_{\mu\nu}\; u(q_2)\cdot(p_1^{\nu}-p_2^{\nu})\;,
\end{equation}
where $D_{\mu\nu}$ denotes the gluon propagator, which in Feynman gauge is simply given by
\begin{equation}
 D_{\mu\nu}=-\frac{g_{\mu\nu}}{k^2} \quad \text{with}\; \; k=q_1+q_2\;.
\end{equation}
Choosing the Feynman gauge and contracting the Lorentz indices expression gives
\begin{equation}
 {\cal{M}}\sim \bar{v}(q_1)\;(\slashed{p_1}-\slashed{p_2})\; u(q_2)\,
\label{amp}
\end{equation}
and after squaring and performing the fermion spin sum one obtains
\begin{equation}
 |{\cal{M}}|^2 \sim \text{tr}(\slashed{q_1}(\slashed{p_1}-\slashed{p_2})\slashed{q_2}(\slashed{p_1}-\slashed{p_2}))\;.
\end{equation}

\FIGURE{
\includegraphics[width=6cm]{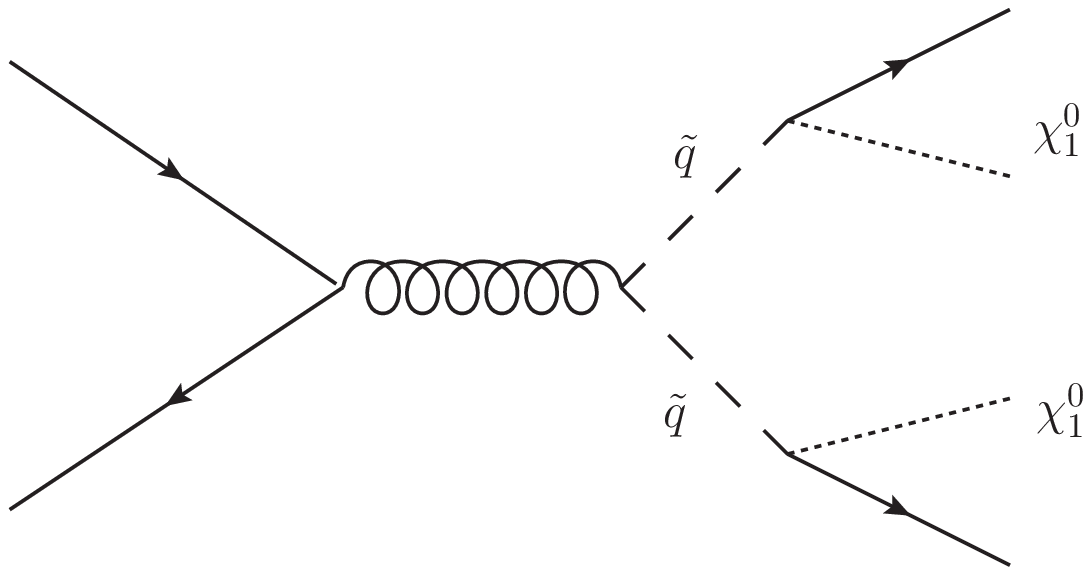}
\caption{Diagram of squark pair production via a s-channel gluon and their subsequent decay.}
\label{squark}
}

As the gluon propagator is gauge dependent, the gauge dependence vanishes only in the sum of all contributing amplitudes.

To calculate the effect of a specific gauge to the given diagram we start with a general covariant gauge.
The gluon propagator can be written as
\begin{equation}
 D_{\mu\nu}=-\frac{1}{k^2}\left(g_{\mu\nu} +(1-\lambda)\frac{k_{\mu}k_{\nu}}{k^2}\right)\;,
\end{equation}
so that, for $\lambda=1$, we recover the Feynman gauge. In the general case the presence of a term $k_{\mu}k_{\nu}$
leads to an extra term in Eq.~(\ref{amp}) of the form $\slashed{k}(k\cdot p_1 -k\cdot p_2)$.
It can easily be seen that this extra term vanishes if one replaces
\begin{equation}
 \slashed{k}=\slashed{q_1}+\slashed{q_2}
\end{equation}
and makes use of the Dirac equation for massless quarks,
\begin{equation}
 \slashed{q_2}\; u(q_2)=0, \quad \bar{v}(q_1)\;\slashed{q_1}=0\;.
\end{equation}
Next, we turn to the case of non-covariant gauges. We consider the following structure for the gluon
propagator:
\begin{equation}
 D_{\mu\nu}=-\frac{1}{k^2}\left(g_{\mu\nu}-\frac{n_{\mu}k_{\nu}+n_{\nu}k_{\mu}}{n\cdot k} +\frac{n^2k_{\mu}k_{\nu}}{(n\cdot k)^2}\right)\;,
\label{noncov}
\end{equation}
where $n$ can be a time-like, space-like or light-like vector.\\
The third term of Eq.~(\ref{noncov}) vanishes with the same argument as for covariant gauges, as well as the term
$\sim n_{\nu}k_{\mu}$. 

The remaining term can be written as
\begin{equation}
 \slashed{n}(k\cdot p_1 -k\cdot p_2)=\slashed{n}(q_1\cdot p_1 +q_2\cdot p_1 -q_1\cdot p_2 -q_2\cdot p_2)\;.
\label{term2}
\end{equation}
Momentum conservation in the on-shell limit implies
\begin{equation}
 q_1\cdot p_1 = q_2\cdot p_2\;, \quad q_2\cdot p_1=q_1\cdot p_2\;,
\end{equation}
and therefore the additional factor in Eq.~(\ref{term2}) is zero.

\providecommand{\href}[2]{#2}\begingroup\raggedright\endgroup

\end{document}